\documentstyle[%
twocolumn,%
aps,%
prl,%
psfig%
]{revtex}

\begin{document}

\preprint{\em preprint}

\draft

\title{Reentrant behavior in the superconducting phase-dependent resistance\\
of a disordered two-dimensional electron gas}

\author{S.~G.~den~Hartog,$^1$  B.~J.~van~Wees,$^1$ Yu.~V.~Nazarov,$^2$ 
T.~M.~Klapwijk,$^1$  and G.~ Borghs,$^3$ }

\address{$^1$Department of Applied Physics and Materials Science Center,
University of Groningen,\\ 
Nijenborgh 4, 9747 AG Groningen, The Netherlands.}

\address{$^2$Faculty of Applied Physics and Delft Institute for 
Microelectronics and Submicron Technology,\\ 
Delft University of Technology,
Lorentzweg 1, 2628 CJ Delft, The Netherlands.}

\address{$^3$Interuniversity Micro Electronics Center, Kapeldreef 75,
B-3030 Leuven, Belgium}

%\date{DRAFT: to be submitted to Phys. Rev. B Rapid Communications}

\maketitle

\begin{abstract}
We have investigated the bias-voltage dependence of the phase-dependent 
differential resistance of a disordered $T$-shaped two-dimensional electron 
gas coupled to two superconducting terminals.
The resistance oscillations first increase upon 
lowering the energy. For bias voltages below the Thouless energy,
the resistance oscillations are suppressed and disappear almost completely 
at zero bias voltage.
We find a qualitative agreement with the calculated reentrant behavior of
the resistance and discuss quantitative deviations.
\end{abstract}

 % PACS numbers

Over the past years experimental and theoretical investigations 
have revealed how the resistance of a normal conductor ($N$) strongly 
coupled to a superconductor ($S$) is modified 
due to the superconducting proximity effect.
At the $NS$ interface electrons are converted in Andreev reflected holes 
incorporating the macroscopic phase of the superconductor. 
Superconducting correlation between electrons and holes penetrates 
a distance $\xi(\epsilon)\!\equiv\! \sqrt{\hbar D/\epsilon}$ 
into the disordered normal conductor with diffusion constant $D$, 
where $\epsilon$ denotes the relevant energy 
(maximum of temperature $k_B T$ or bias voltage $eV$). 
A striking prediction by Artemenko {\em et al.} \cite{Arte79} in 1979  
was that the resistance of a disordered normal conductor at low energies 
($k_BT,eV\rightarrow0$) returns to its full normal-state resistance $R_N$, 
despite the presence of superconducting correlation.
Originally, this prediction was valid for short disordered normal conductors
[length $L\!\ll\!\xi(\Delta_S)$, with superconducting energy gap $\Delta_S$].
Recently, its validity has been extended to long disordered normal conductors
 coupled to superconductors \cite{Naza96,theo1,theo2}.
Theoretical analysis based on impurity-averaged Keldysh Green's function 
techniques \cite{Naza96,theo1} has shown that transport can be described 
by an effective diffusion constant, which depends on energy and position.
At low and high energies this effective diffusion constant returns to $D$,
its value in the normal state, and
for intermediate energies it is enhanced at position $\xi(\epsilon)$.
Therefore, the reduction in resistance should vanish for 
both low and high energies and thus display a {\em reentrant} behavior.
The maximum reduction depends on the particular 
shape of the normal conductor and occurs around an energy of a few times the 
Thouless energy $E_T \!\equiv\! \hbar D/L^2$.  

Several experimental observations regarding the reentrance of the 
resistance have been reported using different geometries of the 
normal conductor.   
We have reported a suppression at low energies of the superconducting 
phase-dependent resistance of a $T$-shaped two-dimensional electron gas (2DEG) 
coupled to two superconductors 
employing an Andreev interferometry technique \cite{Hart96a}. 
In a cross-shaped 2DEG interferometer \cite{Hart96b}, 
we observed a similar behavior.
Charlat {\it et al.} \cite{Char96} have studied the magnetoresistance 
of a normal metal Aharonov-Bohm ring coupled to a single superconductor.
Recently, they have also studied a single $T$-shaped metal wire connected to a 
single superconductor \cite{Char97}.
Finally, Petrashov {\it et al.} \cite{Petr96} have observed 
a reentrant behavior in a cross-shaped normal metal interferometer.
Note, however, that their interpretation is hindered by the presence of a 
circulating supercurrent \cite{Wees96}, which drastically
modifies both the magnitude of the superconducting phase-dependent resistance 
oscillations as well as its shape. 
In Ref.~\cite{Hart96b} we were able to conclude that the effect of a 
circulating supercurrent was negligible, which also holds for the 
$T$-shaped 2DEG interferometer reported here.

So far the zero-energy limit where the resistance should return fully
to $R_N$ has not been reached. 
The experiments \cite{Hart96a,Char97,Petr96} have only reported an increase 
in resistance at low energies of about 55\% of the maximum reduction 
in resistance. 

\begin{figure}
\centerline{\psfig{figure=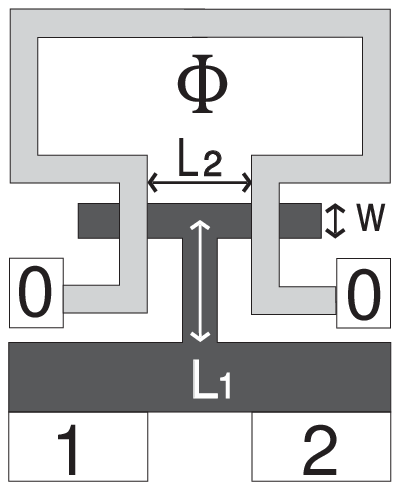,width= 0.185 \textwidth} 
	\psfig{figure=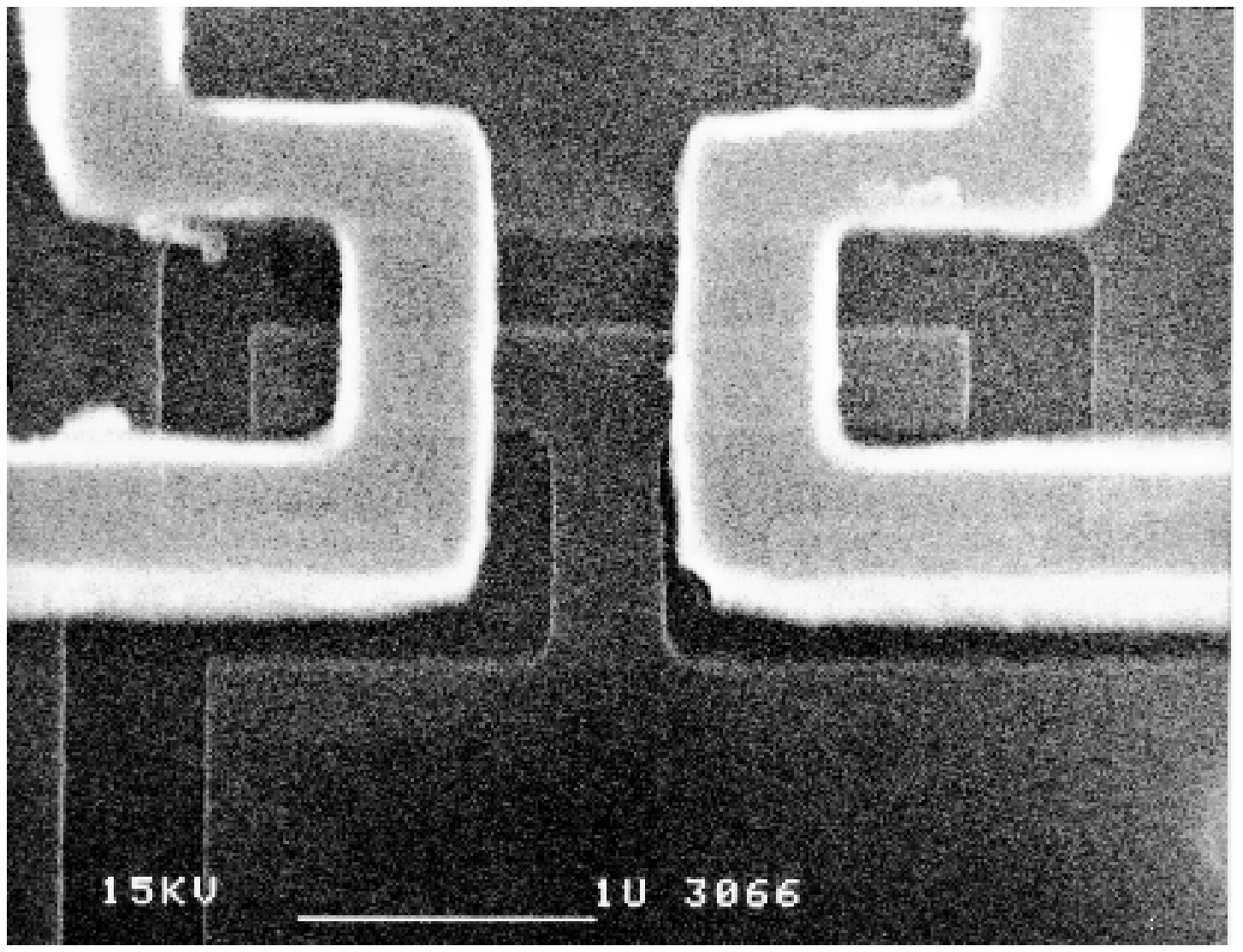,width= 0.30 \textwidth}}
\caption[]{Sample layout. 
The left-hand panel shows a schematic picture of a $T$-shaped 2DEG with an 
interrupted superconducting loop. 
The contacts (0) are connected to the niobium loop and (1) and (2) 
are connected to the $T$-shaped 2DEG. 
%The dimensions are $L_1 \!\simeq\! 0.96~\mu$m, 
%$L_2 \!\simeq\! 0.73~\mu$m and $W \!\simeq\! 0.34~\mu$m.
The right-hand panel shows a scanning electron micrograph of the device.}
\label{layout} 
\end{figure}

In this paper, we will report a reentrant
behavior observed in the bias-voltage dependence of the resistance of a 
$T$-shaped 2DEG coupled to two superconducting terminals.
Around zero energy the resistance oscillations 
due to the superconducting phase are almost completely suppressed,
confirming the theoretical predictions \cite{Arte79,Naza96,theo1,theo2}.  
We will compare our data with theoretically calculated resistances,
which will highlight the consequences of using a low electron-density 
semiconductor instead of a normal metal.

We have reinvestigated the same devices ($A$ and $D$) as studied in 
Ref.~\cite{Hart96a}. Instead of focussing on the sample-specific
resistance oscillations in higher magnetic fields, 
we will focus on the energy dependence of the low magnetic field
resistance oscillations. 
For this purpose, we have included additional filtering in the leads 
connected to the device at cryogenic temperatures \cite{Mart87}. 

Our $NS$ interferometer consists of a $T$-shaped
2DEG attached to two superconducting terminals (see Fig.~\ref{layout}). 
These superconducting terminals (0) are the ends of an
interrupted superconducting loop, which forces the electrochemical 
potential of the superconducting terminals to be equal.
The superconducting phase difference 
$\delta \varphi\!=\!2\pi \Phi/\Phi_0$, with $\Phi_0\!\equiv\!h/2e$, 
between both terminals can be varied linearly by an applied magnetic 
flux $\Phi$ through this interrupted superconducting loop
(area $10.3 \mu\mbox{\rm m}^2$).

The $T$-shaped 2DEG has been formed in an InAs/AlSb heterostructure,
since highly transparent interfaces can be obtained between 
superconductors and the 2DEG in the underlying InAs layer. 
After removing the AlSb top layer,
insulating trenches were defined in the InAs layer by wet chemical etching. 
Subsequently, 50 nm Nb superconducting electrodes were deposited 
after {\it in situ} Ar cleaning of the exposed InAs surface \cite{Magn95}.
The transport properties of the InAs channel are roughly characterized by an
electron density $n_s\!\sim\!1.5\!\times\!\!10^{16} \mbox{\rm m}^{-2}$
and an elastic mean free path $\ell_e\!\sim\!0.2 \mu$m. 
The length of the vertical arm of the T-shaped 2DEG is $L_1\!=\!0.96 \mu$m,
the separation between both superconducting terminals is $L_2\!=\!0.73 \mu$m,
and the width of the horizontal arm is $W\!=\!0.34 \mu$m.

\begin{figure}[thb]
\centerline{\psfig{figure=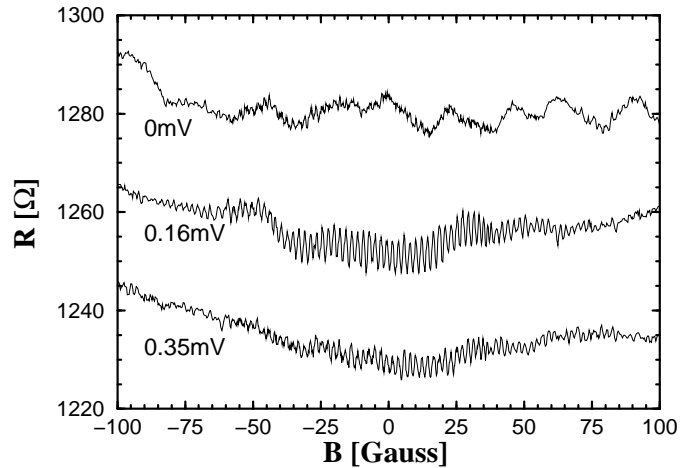,width= 0.5 \textwidth}}
\caption[]{Differential magnetoresistance $R_{01,02}$ of device $A$ at 250 mK 
for three applied dc bias voltages: 0 mV, 0.16 mV, and 0.35 mV 
(from top to bottom). 
These traces are offsetted by 55 G to compensate for the reminant 
magnetic field of the superconducting magnet.}
\label{traces} 
\end{figure}

The differential resistance $R_{01,02}$ is plotted versus applied 
magnetic field in Fig.~\ref{traces}.
The period of the resistance oscillations corresponds 
with the expected flux quantum $\Phi_0$ through the area of the 
interrupted Nb loop.
The magnetic flux is  not only present in this interrupted Nb loop,
but also in the $T$-shaped 2DEG itself. 
The additional phase-shifts due to this magnetic flux are expected to 
destroy superconducting correlations, and, consequently, the 
resistance oscillations when roughly one flux quantum $h/e$ penetrates the 
$T$-shaped 2DEG (area 0.52 $\mu$m$^2$), which corresponds to about 50 G 
including a magnetic flux enlargement of about 1.7 due to the Meissner effect.
For devices $A$ and $D$ the resistance oscillations disappeared around 
respectively 80 and 120 G, or equivalently
an actual magnetic flux of about 1.6 and 2.4 $h/e$.

We have investigated the energy dependence of the phase-dependent differential
resistance $R_{01,02}$ by varying the applied dc bias voltage 
(see Fig.~\ref{traces}). 
Upon decreasing $V_{dc}$ from 0.35 mV to 0.16 mV
the magnitude of the resistance oscillations increases.
The magnetoresistance oscillations plotted in the upper trace are almost
disappeared, which was recorded when all relevant energies were 
smaller than $E_T$: $V_{dc}$=0 mV and $eV_{ac}\!<\!k_B T\!\simeq$0.02 meV.

\begin{figure}[thb]
\centerline{\psfig{figure=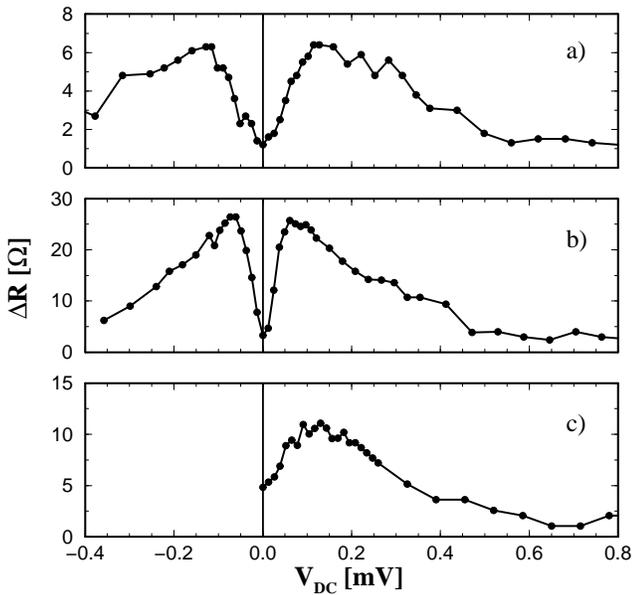,width= 0.5 \textwidth}}
\caption[]{Energy dependence of the top-top magnitude of the resistance 
oscillations: 
(a) device $A$ at 250 mK with cryogenic filtering 
($R_N\!\simeq\!1250~\Omega$), 
(b) device $D$ at 170 mK with cryogenic filtering 
($R_N\!\simeq\!1200~\Omega$), 
and (c) device $D$ at 50 mK without cryogenic filtering   
($R_N\!\simeq\!1450~\Omega$, reproduced from Ref.~\cite{Hart96a}).}
\label{overview}
\end{figure}

The complete energy dependence of the magnitude of these resistance 
oscillations for device $A$ is collected in Fig.~\ref{overview} (a).
Note that the bias-voltage dependence directly reflects the 
energy dependence, whereas the temperature dependence corresponds to a 
convolution of the bias-voltage dependence with the Fermi-Dirac distribution.
The oscillations appear below a dc bias voltage of about $\pm0.5$ mV, 
which is well below the superconducting energy gap $\Delta_S$ of 1.3 mV.
The resistance oscillations reach a maximum magnitude around 0.1 mV, 
which is suppressed by about 80\% at zero bias voltage.
For comparison, we have plotted the bias-voltage dependence for the other 
device $D$ in Figs.~\ref{overview} (b) and \ref{overview} (c).
The data of Fig.~\ref{overview} (c) are copied from  Ref.~\cite{Hart96a} 
and were obtained without using cryogenic filtering.
The main difference is visible around zero bias voltage,
where the resistance oscillations are less suppressed due to
an elevated noise temperature.
However, the energy dependence of the resistance oscillations is
qualitatively identical for both devices and displays the predicted
reentrant behavior in the resistance.

\begin{figure}[thb]
\centerline{\psfig{figure=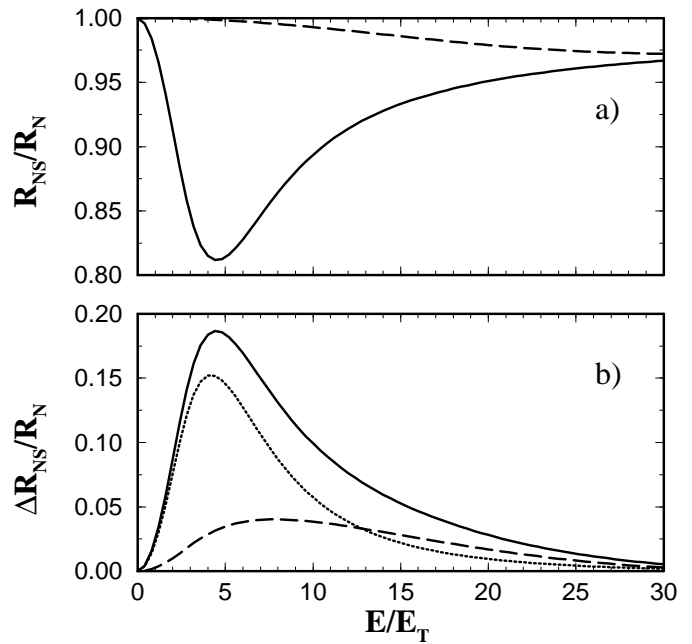,width= 0.5 \textwidth}}
\caption[]{Calculated energy dependence of the resistance for a $T$-shaped
normal conductor with $L_1$=$L_2$=$L$, where $E_T$=$\hbar D/L^2$.
Note that the temperature dependence can be obtained by convoluting this
energy dependence with the Fermi-Dirac distribution.
(a) displays the resistance at $\delta \varphi=0$ (solid line)
and at $\delta \varphi\!=\!\pi$ (dashed line).
(b) displays the magnitude of the total resistance oscillations 
(solid line), which is the sum of the resistance oscillations of the vertical 
arm (dotted line) and both horizontal arms in parallel (dashed line).}
\label{calc}
\end{figure}

The calculated energy dependence of the resistance $R_{01,02}$
for a $T$-shaped interferometer is shown in Fig.~\ref{calc}.
The length of the vertical and horizontal arms was assumed to be the same 
($L_1$=$L_2$=$L$).
The procedure is based on evaluating nonequilibrium 
quasiclassical Green's functions for diffusive superconductors 
using the Keldysh technique.
For detailed information about the calculations we refer to Ref.~\cite{Naza96}.
Here, we will only emphasize the assumptions which are made.
First, the effective diffusion coefficient varies only in one dimension 
(along the wire length), transport is fully diffusive ($L\gg\ell_e$), 
and phase-breaking events are neglected ($L\ll \ell_\phi$).
Second, the energy-dependent phase-shift during Andreev reflection
is disregarded, which implies that all relevant energies  
($k_B T, eV_{dc},E_T$) are assumed to be small compared to $\Delta_S$.
Third, the pair potential $\Delta_N$ in the normal conductor is neglected, 
which means that the electron-electron interaction strength is assumed 
to be zero.

In Fig.~\ref{overview} (a) the resistance at $\delta\varphi$=0 (solid line) 
describes the reentrant behavior of the resistance in zero magnetic field. 
The total resistance comprises the resistance of the vertical arm 
and the resistances of both horizontal arms in parallel. 
The small reduction in resistance at $\delta\varphi$=$\pi$ (dashed line)
below $R_N$ solely arises from the horizontal arms, 
which disappears when $L_2\!\ll\!L_1$.
In Fig.~\ref{overview} (b) the magnitude of the resistance oscillations is 
plotted, which is the difference between the two curves of 
Fig.~(\ref{overview} (a).
Figure~\ref{overview} (b) also shows that the contribution of the 
vertical arm dominates over the contribution of 
both horizontal arms in parallel.
The resistance oscillations are thus expected to have a maximum 
magnitude of about 18\%$R_N$ around an energy $eV\!\simeq\!4.5E_T$.

Let us now proceed with a quantitative comparison between experiment 
and theory.
The Thouless energy corresponding with the vertical arm of the $T$-shaped
2DEG amounts to $E_T$=$\hbar D/L_1^2\!\simeq$0.06 meV for both devices,
since the normal-state resistances are roughly identical.
For devices $A$ and $D$ the maximum magnitude of the resistance oscillations 
are, respectively, about 0.5\% and 2.2\% of $R_N$ at an energy of 
2.4$E_T$ and 1.2$E_T$.  
Although two nominally identical devices show a variation in
both energy dependence and magnitude of the resistance oscillations,
they do differ significantly from the theoretical expectation.
In principle a nonideal $NS$-interface could be responsible for
a reduction in magnitude. 
However, the $NS$ interface resistance is small compared to the diffusive 
resistance of the $T$-shaped 2DEG. 
In general the resistance at $\delta\varphi$=0 is reduced below $R_N$ 
for $eV\!<\!\Delta_S$, which implies
that the probability for Andreev reflection dominates over normal reflection.

When we correct for the finite temperature in this experiment,
the estimated magnitude of the resistance oscillations at zero bias voltage 
for device A will be slightly reduced and for device D  
will become negative.
This nonzero magnitude at zero temperature and bias voltage results
from the presence of sample-specific conductance fluctuations 
modulated by the superconducting phase \cite{Hart96a,Hart96b}.
These oscillations are present at all energies and magnetic fields.
Their rms magnitude is about 1 $\Omega$ for device $A$ and 2 $\Omega$ for 
device $D$.
The magnitude of the observed resistance oscillations at zero bias voltage 
is for both devices suppressed to this sample-specific magnitude.
Therefore, we have confirmed the theoretical prediction 
\cite{Arte79,Naza96,theo1,theo2} that the nonsample-specific phase-dependent 
resistance vanishes at zero energy.
Note that we have also reinvestigated the reentrant behavior in the 
two-terminal resistance of the cross-shaped 2DEG interferometer \cite{Hart96b} 
using cryogenic filtering. Here, the oscillation magnitude 
around zero bias voltage was reduced by a factor of 2, 
which was limited by an enhanced magnitude of sample-specific 
oscillations due to an higher resistance of these devices.

Another contribution to the resistance oscillations at zero bias voltage 
could originate from the fact that our devices
are not precisely diffusive, since $L\!\simeq\!5\ell_e$.
We can correct for that in the calculation by
including a quantum point contact (QPC) with a resistance of about 
350$\Omega$ in front of diffusive resistors modeling 
the $T$-shaped 2DEG \cite{Been94}.
Note that this QPC resistance does {\em not} exhibit a reentrant behavior
and is predicted to show phase-induced oscillations at zero energy 
\cite{Been94}.
We have calculated that for our geometry the expected magnitude for these 
QPC resistance oscillations should be less than 10\% of the maximum 
resistance oscillations around 2.7$E_T$.
Therefore, we do not believe that for this device these QPC resistance 
oscillations could be responsible for a significant contribution 
to the observed oscillations around zero bias voltage \cite{Hart97}.

So far we did not mention the energy dependence of the 
resistance at $\varphi$=0.
Figure~\ref{traces} shows that an increase in applied bias voltage
causes a decrease in resistance. 
The resistance well above $E_T$ (bottom trace) does not show an 
increase, which seems to be in contrast with the reentrant behavior.
However, a similar-sized 2DEG wire without superconducting terminals
also shows an increase in resistance around zero bias voltage \cite{Hart98},
which was attributed to be the Coulomb anomaly in the resistance 
caused by electron-electron interactions (EEI's).
Apparently, the contribution to the resistance caused by EEI's 
masks the reentrant behavior in the resistance.
An interesting remark is that EEI's can give rise to a finite superconducting 
pair potential $\Delta_N$ in the normal conductor, which can be modulated
by the superconducting phase \cite{Naza96}. 
The resulting magnitude of the resistance oscillations depends
on the strength of EEI's and is in general much smaller compared to the
reentrant behavior in the resistance.
In this experiment, the magnitude of these resistance oscillations caused by 
EEI's is smaller than the magnitude of the sample-specific oscillations.
 
The last issue we adress is the shape of the resistance oscillations,
which for a slightly different geometry was predicted to develop a 
strong nonsinusoidal shape \cite{Naza96}.
However, we observe only a very small contribution of higher harmonics.
Around energies where the total magnitude exhibits a maximum
the magnitude of the first higher harmonic (period $h/4e$) is about 
1.5 $\Omega$ for device $D$.
For device $A$ we could not detect the second harmonic. 
Note that also  Petrashov {\em et al.} \cite{Petr96} observed that the 
resistance oscillations are sinusoidal after correcting for the 
extrinsic deformation caused by the circulating supercurrent. 
Presently, we do not have a good understanding why certain geometries
favor sinusoidal oscillations.

In conclusion, we have investigated  in detail the reentrant behavior of the 
superconducting phase-dependent resistance of a $T$-shaped 2DEG interferometer.
The magnitude of the resistance oscillations at zero energy  
was shown to be suppressed to the magnitude of the phase-dependent 
sample-specific conductance fluctuations. 
The shape of the phase-dependent resistance strongly 
deviates from the theoretical predictions, which remains an open issue.  
\vspace{\baselineskip}    

This work was part of the research program of the stichting voor
Fundamenteel Onderzoek der Materie (FOM), which was financially supported
by the Nederlandse organisatie voor Wetenschappelijk Onderzoek (NWO).
B.J. van Wees acknowledges support from the Royal Dutch Academy of
Sciences (KNAW).

\bibliographystyle{prsty}

\end{document}